\documentclass[machines,article,accept,pdftex,moreauthors]{Definitions/mdpi} 
\usepackage{algorithmic}
\usepackage{algorithm}
\usepackage{mathtools}
\usepackage{graphicx}
\usepackage{bm}

\firstpage{1} 
\pubvolume{1}
\issuenum{1}
\articlenumber{0}
\pubyear{2024}
\copyrightyear{2024}
\externaleditor{Academic Editor: Yahui Liu}
\datereceived{7 March 2024} 
\daterevised{1 April 2024} 
\dateaccepted{15 April 2024 } 
\datepublished{ } 
\hreflink{https://doi.org/} 

\Title{Autonomous Vehicle Decision and Control through Reinforcement Learning with Traffic Flow Randomization}

\TitleCitation{Autonomous Vehicle Decision and Control through Reinforcement Learning with Traffic Flow Randomization}

\Author{Yuan Lin 
 \orcidB{}, Antai Xie *\orcidA{} and Xiao Liu}

\AuthorNames{Yuan Lin, Antai Xie and Xiao Liu}

\AuthorCitation{Lin, Y.; Xie, A.; Liu, X.}

\address[1]{Shien-Ming Wu School of Intelligent Engineering, South China University of Technology,\linebreak Guangzhou 510641, China; yuanlin@scut.edu.cn (Y.L.); 202121060431@mail.scut.edu.cn (X.L.) 
}

\corres{\hangafter=1 \hangindent=1.05em \hspace{-0.82em}Correspondence: 202220159307@mail.scut.edu.cn}

\abstract{Most of the current studies on autonomous vehicle decision-making and control based on reinforcement learning are conducted in simulated environments. The training and testing of these studies are carried out under the condition of rule-based microscopic traffic flow, with little consideration regarding migrating them to real or near-real environments. This may lead to performance degradation when the trained model is tested in more realistic traffic scenes. In this study, we propose a method to randomize the driving behavior of surrounding vehicles by randomizing certain parameters of the car-following and lane-changing models of rule-based microscopic traffic flow. We trained policies with deep reinforcement learning algorithms under the domain-randomized rule-based microscopic traffic flow in freeway and merging scenes and then tested them separately in rule-based and high-fidelity microscopic traffic flows. The results indicate that the policies trained under domain-randomized traffic flow have significantly better success rates and episodic rewards compared to those trained under non-randomized traffic flow.}


\begin{document}
\section{Introduction}
In recent years, autonomous vehicles have received increasing attention as they have the potential to free drivers from the fatigue of driving and facilitate efficient road traffic~\cite{le2015autonomous}. With the development of machine learning, rapid progress has been achieved in the development of autonomous vehicles. In particular, reinforcement learning, which enables vehicles to learn driving tasks through trial and error, continuously improves the learned policies. Compared to supervised learning, reinforcement learning does not require the manual labeling or supervision of sample data~\cite{sallab2017deep,hoel2018automated,ye2019automated,ye2020automated}. However, reinforcement learning models require tens of thousands of trial-and-error iterations for policy learning, and real vehicles on the road can hardly withstand so many trials. Therefore, the current mainstream research on autonomous driving with reinforcement learning focuses on using virtual driving simulators for training.

Lin et al.~\cite{lin2020anti} utilized deep reinforcement learning within a driving simulator, Simulation of Urban Mobility (SUMO), to train autonomous vehicles, enabling them to merge safely and smoothly at on-ramps. Peng et al.~\cite{peng2022integrated} also employed deep reinforcement learning algorithms within a SUMO to train a model for lane changing and car following. They tested the model by reconstructing scenes using NGSIM data, and the results indicate that the models based on reinforcement learning demonstrate higher efficacy than those based on rule-based approaches. Mirchevska et al.~\cite{mirchevska2017reinforcement} used fitted Q-learning for high-level decisionmaking on a busy simulated highway. However, the microscopic traffic flows of these studies are based on rule-based models, such as the Intelligent Driver Model (IDM)~\cite{treiber2000congested,liu2021deep,huang2022tip} and the Minimize Overall Braking Induced by Lane Change (MOBIL) model. These are mathematical models based on traffic flow theory~\cite{punzo2005analysis}. They tend to simplify vehicle motion behavior and do not consider the interaction of multiple vehicles. Autonomous vehicles trained with reinforcement learning in such microscopic traffic flows may perform exceptionally well when tested in the same environments. However, when the trained model is applied to more realistic or real-world traffic flows, their performance may significantly deteriorate, and they could even cause traffic accidents. This is due to the discrepancies between simulated and real-world traffic flows.

For research on sim-to-real transfer, numerous methods have been proposed to date. For instance, robust reinforcement learning has been explored to develop strategies that account for the mismatch between simulated and real-world scenes~\cite{tessler2019action}. Meta-learning is another approach that seeks to learn adaptability to potential test tasks from multiple training tasks~\cite{wang2016learning}. Additionally, the domain randomization method used in this article is acknowledged as one of the most extensively used techniques to improve the adaptability to real-world scenes~\cite{andrychowicz2020learning}. Domain randomization relies on randomized parameters aimed at encompassing the true distribution of real-world data. Sheckells et al.~\cite{sheckells2019using} applied domain randomization to vehicle dynamics, using stochastic dynamic models to optimize the control strategies for vehicles maneuvering on elliptical tracks. Real-world experiments indicated that the strategy was able to maintain performance levels similar to those achieved in simulations. However, few studies have applied domain randomization to microscopic traffic flows and investigated its efficacy.

In recent years, many driving simulators have been moving towards more realistic scenes. One type includes data-based driving simulators (InterSim~\cite{sun2022intersim} and TrafficGen~\cite{feng2023trafficgen}), which train neural network models by extracting vehicle motion characteristics from real-world traffic datasets, resulting in interactive microscopic traffic flows. However, the simulation time is much longer than for most rule-based driving simulators due to the complexity of the models. The other kind includes theory-based interactive traffic simulators, which can generate long-term interactive high-fidelity traffic flows by combining multiple modules (LimSim~\cite{wenl2023limsim}). The traffic flow generated by LimSim closely resembles an actual dataset with a normal distribution, sharing similar means and standard deviations~\cite{zheng2022citysim}.

This paper proposes a domain randomization method for rule-based microscopic traffic flows for reinforcement learning-based decision and control. The parameters of the car-following and lane-changing models are randomized with Gaussian distributions, making the microscopic traffic flows more random and behaviorally uncertain, thus exposing the agent to a more complex and variable driving environment during training. To investigate the impact of domain randomization, this paper will train and test agents using microscopic traffic flow without randomization, high-fidelity microscopic traffic flow, and domain-randomized traffic flow for freeway and merging scenes.

The rest of this paper is structured as follows: Section \ref{sec:Microscopic Traffic Flow} introduces the relevant microscopic traffic flows. Section \ref{sec:Domain Randomization Microscopic Traffic Flow} describes the proposed domain randomization method. Section \ref{sec:Simulation Experiment} presents the simulation experiments and the analysis of the results for the freeway and merging scenes. Finally, the conclusions are drawn in Section \ref{sec:Conclusion}.

\section{Microscopic Traffic Flow}\label{sec:Microscopic Traffic Flow}
Microscopic traffic flow models take individual vehicles as the research subject and mathematically describe the driving behaviors of the vehicles, such as acceleration, overtaking, and lane changing.
\subsection{Rule-Based Microscopic Traffic Flow}

This paper utilizes IDM and SL2015 as the default car-following and lane-changing models, respectively. The following is a detailed introduction to them.
\subsubsection{IDM Car-Following Model}
IDM was originally proposed by Treiber in~\cite{treiber2000congested}, capable of describing various traffic states from free flow to complete congestion with a unified formulaic approach. The model takes the preceding vehicle's speed, the ego vehicle's speed, and the distance to the preceding vehicle as inputs to output the ego vehicle's safe acceleration. The acceleration of the ego vehicle at each timestep is 

\begin{equation}\label{eq:idm_v}
\dot{v}(t) = a \left[1 - \left(\frac{v(t)}{v_0}\right)^\delta - \left(\frac{s^*(v(t), \Delta v(t))}{s}\right)^2\right],
\end{equation}
where $a$ represents the maximum acceleration of the ego vehicle, $v(t)$ is the current speed of the ego vehicle, $v_0$ is the desired speed of the ego vehicle, $\delta$ is the acceleration exponent, $\Delta v(t)$ is the speed difference between the ego vehicle and the preceding vehicle, $s$ is the current distance between the ego vehicle and the preceding vehicle, and $s^*(v(t), \Delta v(t))$ is the desired following distance. The desired distance is defined as follows:
\begin{equation}\label{eq:idm_s}
s^*(v(t), \Delta v(t)) = s_0 + \max\left(0, v(t) * T + \frac{v(t) * \Delta v(t)}{2\sqrt{ab}}\right),
\end{equation}
where $s_0$ is the minimum gap, $T$ is the bumper-to-bumper time gap, and $b$ represents the maximum deceleration.
\subsubsection{SL2015 Lane-Changing Model}\label{sec:sl2015}
The safety distance required for the lane-changing process is calculated as follows:
\begin{equation}\label{eq:sl2015}
d_{\text{lc,veh}}(t) = 
\left\{
\begin{array}{ll}
v(t) * a_1 + 2l_{\text{veh}} , & \text{if } v(t) \leq v_{\text{c}}, \\
v(t) * a_2 + 2l_{\text{veh}} , & \text{if } v(t) > v_{\text{c}},
\end{array}
\right.
\end{equation}
where \(d_{\text{lc,veh}}(t)\) denotes the safety distance required for lane changing, \(v(t)\) represents the velocity of the vehicle at time \(t\), \(l_{\text{veh}}\) is the length of the vehicle, \(a_1\) and \(a_2\) are safety factors, and the threshold speed \(v_{\text{c}}\) differentiates between urban roads and highways. 

The profit \( b_{ln}(t) \) at time \( t \) for changing lanes is calculated as follows:
\begin{equation}
b_{ln}(t) = \frac{v(t, ln) - v(t, lc)}{v_{\text{max}}(lc)},
\end{equation}
where \( v(t, ln) \) is the velocity of the vehicle in the target lane at the next timestep, \( v(t, lc) \) is the safe velocity in the current lane, and \( v_{\text{max}}(lc) \) is the maximum velocity allowed in the current lane. The goal here is to maximize the velocity difference, thereby increasing the benefit of changing lanes.

If the profit $b_{ln}(t)$ for the current timestep is greater than zero, then this profit will be added to the cumulative profit. Conversely, if the profit for the current timestep is less than zero, the cumulative profit will be halved to moderate the desire to change to the target lane. If the cumulative profit is larger than a threshold, lane change can be initiated.

\subsection{LimSim High-Fidelity Microscopic Traffic Flow}
The study employs the LimSim driving simulation platform's high-fidelity microscopic traffic flow. The high-fidelity microscopic traffic flow in LimSim is based on optimal trajectory in the Frenet frame~\cite{werling2010optimal}. Within the circular area around the ego vehicle, the microscopic traffic flow is updated based on each optimal trajectory.
\subsubsection{Trajectory Generation}
\label{sec:generate}
In the Frenet coordinate system, the motion state of a vehicle can be described by the tuple \([s, \dot{s}, \ddot{s}, d, \dot{d}, \ddot{d}]\), where \(s\) represents the longitudinal displacement, \(\dot{s}\) the longitudinal velocity, \(\ddot{s}\) the lateral acceleration, \(d\) represents the lateral displacement, \(\dot{d}\) the lateral velocity, and \(\ddot{d}\) the latitudinal acceleration. 

\subsubsection*{Lateral Trajectory Generation 
}

The lateral trajectory curve can be expressed by the following fifth-order polynomial:
\begin{equation}
d(t) = a_{d0} + a_{d1}t + a_{d2}t^2 + a_{d3}t^3 + a_{d4}t^4 + a_{d5}t^5.
\end{equation}
The trajectory start point is known as \(D_0 = [d_0, \dot{d}_0, \ddot{d}_0]\), and a complete polynomial trajectory can be determined once the end point \(D_1 = [d_1, \dot{d}_1, \ddot{d}_1]\) is specified. As vehicles travel on the road, they use the road centerline as the reference line for navigation, and the optimal state should be moving parallel to the centerline, which means the end point would be \(D_1 = [d_1, 0, 0]\). Equidistant sampling points are selected between the start point and end point, and the multiple polynomial segments are connected to form many complete lateral~trajectories.

\subsubsection*{Longitudinal Trajectory Generation}
Longitudinal trajectory curve can be expressed with a fourth-degree polynomial:
\begin{equation}
s(t) = a_{s0} + a_{s1}t + a_{s2}t^2 + a_{s3}t^3 + a_{s4}t^4 .
\end{equation}
\( S_0 = [s_0, \dot{s}_0, \ddot{s}_0] \) is the start point and \( S_1 = [\dot{s}_1, \ddot{s}_1] \) is the end point. Equidistant sampling points are selected between the start point and end point, and the multiple polynomial segments are connected to form many complete longitudinal trajectories.
\subsubsection{Optimal Trajectory Selection}
\label{sec:selection}
The trajectory selection process involves evaluating a cost function that includes key components: trajectory smoothness, which is determined by the heading and curvature differences between the actual and reference trajectories; vehicle stability, indicated by the differences in acceleration and jerk between the actual and reference trajectories; collision risk, assessed by the risk level of collision with surrounding vehicles; speed deviation, gauged by the velocity difference between the actual trajectory and the reference speed; and lateral trajectory deviation, measured by the lateral distance difference between the actual trajectory and the reference trajectory.

\textls[15]{The total cost function is utilized to evaluate the set of candidate trajectories in Section~\ref{sec:generate}, followed by an assessment of their compliance with vehicle dynamics constraints, such as turning radius and speed/acceleration limits. The trajectory that not only satisfies the vehicle dynamics constraints but also incurs the minimum cost is selected as the final valid trajectory.}

Vehicles within a 50 m perception range of the ego vehicle will be subject to the Frenet optimal trajectory control, with a trajectory being planned every 0.5 s and having a duration of 5 s.

\section{Domain Randomization for Rule-Based Microscopic Traffic Flow}\label{sec:Domain Randomization Microscopic Traffic Flow}
The domain randomization method is based on randomizing the model parameters in the IDM car-following model and the SL2015 lane-changing model. The randomized parameters are shown in Table \ref{t:domain} and are described below.

There are five randomized parameters in the IDM model. ``\(\delta\)'' is the acceleration exponent and ``\(T\)'' is the time gap in the IDM model, respectively. ``$a_{max}$''
, ``\(a_{min}\)'', and ``\(v_{max}\)'' are the upper and lower limits of vehicle acceleration and the upper limit of vehicle speed,~respectively. 

There are two randomized parameters in the SL2015 model. ``\texttt{lcSpeedGain
}'' indicates the degree to which a vehicle is eager to change lanes to gain speed; the larger the value, the more inclined the vehicle is to change lanes. ``\texttt{lcAssertive}'' is another parameter that significantly influences the driver's lane-changing model~\cite{berrazouane2019analysis}; a lower ``\texttt{lcAssertive}'' value makes the vehicle more inclined to accept smaller lane-changing gaps, leading to more aggressive lane-changing behavior. 

Ref.~\cite{kusari2022enhancing} found that the parameters $\delta$, $T$, $a_{max}$, $a_{min}$, and $v_{max}$ are close to Gaussian distributions. Consequently, we adopt Gaussian distributions for all the domain-randomized parameters. All the randomized parameters follow Gaussian distributions within the interval $[s_{min}, s_{max}]$, with distribution \(s\) \((\mu, \sigma^2)\). Here, $s_{max}$ and $s_{min}$ are the upper and lower bounds of the randomization interval. \(\mu\) is set to be \((s_{max} + s_{min}) / 2\), and \(\sigma\) is set to be \((s_{max} - s_{min}) / 6\). Thus, when a vehicle is generated, the probability that its randomized parameter value will fall within \([s_{min}, s_{max}]\) is 99.73\%.

When each vehicle is initialized on the road for each episode, these randomized parameters are generated and assigned to it.
\begin{table}[H] 
\caption{Domain randomization parameters.\label{tab1}}
\begin{tabularx}{\textwidth}{CCC}
\toprule
\textbf{Parameter}	& \textbf{Default Value} & 
\textbf{Randomization Interval}  \\

\midrule
$\delta$	& $4$ & $[3.5, 4.5]$\\
$T$	& 1~s 
 & $[0.5, 1.5]$~s\\
$a_{max}$	& 2.6~m/s$^2$ & $[1.8, 3.4]$~m/s$^2$\\
$a_{min}$	& $-$4.5~m/s$^2$ & $[-5.5, -3.5]$~m/s$^2$\\
$v_{max}$	& 8.33~m/s & $[7.33, 9.33]$~m/s\\
\texttt{lcSpeedGain}	& $1$ & $[0, 100]$\\
\texttt{lcAssertive}		& $1$ & $[1, 5]$\\
\bottomrule
\end{tabularx}
\label{t:domain}
\end{table}

\section{Simulation Experiment}\label{sec:Simulation Experiment}
In this section, we create freeway and merging environments in the open-source SUMO driving simulator~\cite{krajzewicz2012recent} and establish the communication between SUMO and the reinforcement learning algorithm via TraCI~\cite{wegener2008traci}. The timestep for the agent to select actions and observe  environment state is set at 0.1 s. We create non-randomized microscopic traffic flow, the high-fidelity microscopic traffic flow of LimSim, and the domain-randomized microscopic traffic flows. We train the reinforcement learning-based autonomous vehicles under different microscopic traffic flows in freeway and merging scenes, respectively.

\subsection{Merging} \label{sec:merging}
\subsubsection{Merging Environment}
\begin{figure}[H]
\includegraphics[width=0.45\textwidth]{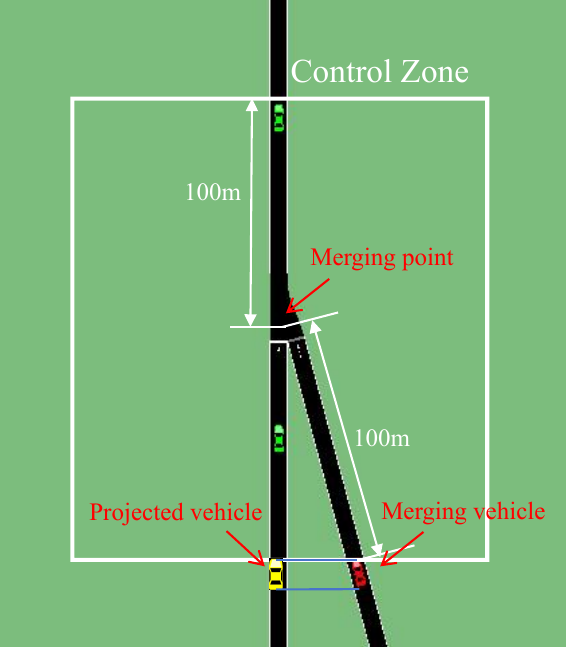}
\caption{Merging in SUMO.}
\label{f:SUMO_merging}
\end{figure}

 We establish the merging environment inspired by Lin et al.~\cite{lin2020anti}. A control zone for the merging vehicle is established, spanning 100 m to the rear of the on-ramp's merging point and 100 m to the front of the merging point, as depicted in Figure \ref{f:SUMO_merging}. The red vehicle, operating under reinforcement learning control, is tasked with executing smooth and safe merging within the designated control area.
 \vspace{-3pt}
\subsubsection*{State}
In defining the state of the reinforcement learning environment, the merging vehicle is projected onto the main road to produce the projected vehicle, and then a total of five vehicles are considered: two vehicles before the projected vehicle, two vehicles after the projected vehicle, and the projected vehicle. In order to utilize the observable information  reasonably, the distance ($d^{p2}_t,d^{p1}_t,d^{f1}_t,d^{f2}_t,d^{m}_t$) of these five vehicles to the merging point, as well as their velocities ($v^{p2}_t,v^{p1}_t,v^{f1}_t,v^{f2}_t,v^{m}_t$), are included in the state representation. These parameters form a state representation with eleven variables, defined as
\begin{equation}\label{eq:merging_s}
s_t = [d^{p2}_t, v^{p2}_t, d^{p1}_t, v^{p1}_t, d^{m}_t, v^{m}_t, a^{m}_t, d^{f1}_t, v^{f1}_t, d^{f2}_t, v^{f2}_t]\in S.
\end{equation}
\subsubsection*{Action}
The action space we have defined is a continuous variable: acceleration within $[-4.5, 2.5]$~m/s$^2$. This range is consistent with the normal acceleration range of surrounding~vehicles.
\begin{equation}
	a_t = \{{acc^{m}_t}\} \in A.
\end{equation}
\subsubsection*{Reward}
We aim for the merging vehicle to maintain a safe distance from the preceding and following vehicles, ensure comfort, and avoid coming to a stop or making the following vehicle brake sharply. Therefore, the reward function is expressed as follows:
\begin{equation}
R_{\text{total}}=R_m + R_b + R_j + R_{\text{stop}} + R_{\text{success}} + R_{\text{collision}}.
\end{equation}
After merging, 
the merging vehicle is safer when its position is in the middle between the preceding and following vehicles. The corresponding penalizing reward is defined as
\begin{equation}
R_m = w_m *  \left( |w| + \frac{\left| \displaystyle \frac{(v_{p1} + v_{f1})}{2} - v_m \right|}{\Delta v_{\text{max}}} \right),
\end{equation}
where $w_m$ represents the weight factor, and $\Delta v_{\text{max}}$ is the maximum allowable speed difference. The variable $w$ is defined to measure the distance  gap among the merging vehicle, its first preceding vehicle, and its first following vehicle. The details are as follows:
\begin{equation}\label{eq:merging_w}
w = \frac{\left| d_{p1} - d_{m} - l_{p1} \right| - \left| d_{m} - d_{f1} - l_{m} \right|}{\left| d_{p1} - d_{f1} - l_{p1} - l_{m} \right|},
\end{equation}
where $l_{p1}$ and $l_{m}$ represent the lengths of the first preceding vehicle and the merging vehicle, both measuring at 5 m. When the first following vehicle performs braking  in the control zone, a penalizing reward is defined as
\begin{equation}
R_b = w_b \ast \frac{\left| a_{f1} \right|}{\max( \left| a_{\text{min}} \right|, a_{\text{max}})},
\end{equation}
where $w_b$ is the weight and $a_{f1}$ is acceleration of the first following vehicle. In order to improve the comfort level of the merging vehicle, we define a penalizing reward for jerk:
\begin{equation}
R_j = w_j \ast \frac{\left| j_m \right|}{j_{\text{max}}} = -w_j \ast \frac{\left| \dot{a}_m \right|}{j_{\text{max}}},
\end{equation}
where $w_j$ is the weight, $j_{\text{max}}$ is maximum allowed jerk, and $\dot{a}_m$ is jerk of the merging vehicle.

In addition, if the merging vehicle comes to a  stop, a penalty of \(R_{\text{stop}}=-0.5\) is imposed. When a merged vehicle collides with any vehicle, a penalty of \(R_{\text{collision}}=-1\) is applied. Conversely, if the merging vehicle successfully reaches its destination, a reward of \(R_{\text{success}}=1\) is granted. Table \ref{t:merging_parameters} shows the values of the above-mentioned parameters of the merging vehicle.
\begin{table}[H]
\caption{Parameter values for the merging vehicle.}
\newcolumntype{C}{>{\centering\arraybackslash}X}
\begin{tabularx}{\textwidth}{lC}
\toprule
\textbf{Parameter} & \textbf{Value} \\
\midrule
Weight for merging midway $w_{m}$ & $-0.015$ \\
Weight for penalizing first following's braking $w_{b}$ & $-0.015$ \\
Weight for penalizing jerk $w_{j}$ & $-0.015$ \\
Maximum allowed speed difference $\Delta v_{\text{max}}$ & 5~m/s \\
Maximum allowed jerk value $j_{\text{max}}$ & 3~m/s$^3$ \\
\bottomrule
\end{tabularx}
\label{t:merging_parameters}
\end{table}

\subsubsection{Soft Actor--Critic}\label{subsec:sac}
SAC is the reinforcement learning algorithm used for training in merging scenes~\cite{haarnoja2018soft}. The SAC algorithm uses the classical framework of reinforcement learning, actor--critic, which helps to optimize the value function and the policy at the same time, and it consists of a parameterized soft-Q function $Q_\theta(s_t, a_t)$ 
and a tractable policy $\pi_\phi(a_t|s_t)$. The parameters of these networks are $\theta$ and $\phi$.
This approach considers a more general maximum entropy objective that not only seeks to maximize rewards but also maintains a degree of randomness in action selection, as follows:
\begin{equation}
	J(\pi) = \sum_{t=0}^{T} \mathbb{E}_{( s_t,  a_t) \sim \rho_\pi} \gamma^t \left[ r( s_t,  a_t) + \alpha H(\pi(\cdot | s_t)) \right],
\end{equation}
where $\rho_\pi$ denotes the state--action distribution under the policy $\pi$, while $H(\pi(\cdot | s_t))$ signifies the entropy of the policy at state $s_t$, thereby enhancing the unpredictability of the chosen actions. The temperature parameter $\alpha$ plays a pivotal role as it calibrates the balance between entropy and reward within the objective function, subsequently influencing the formulation of the optimal policy. The hyperparameters of SAC are the same as in Ref.~\cite{lin2020anti}.


\subsubsection{Results under Different Microscopic Traffic Flows}
\subsubsection*{Training}
In the merging environment, we trained 200,000 timesteps in each of three different microscopic traffic flows. The training was carried out on an NVIDIA RTX 3060 
graphics card paired with an Intel i7-12700F processor. It required approximately 1 h to complete the 
training using both SUMO’s default non-randomized and domain-randomized traffic flows. In contrast, the training under the condition of high-fidelity traffic flow took 3.5 h. Vehicle generation probability is 0.56, and the traffic density on the main road was approximately 16 vehicles per kilometer.

\subsubsection*{Testing}
The trained policy was tested with 1000 episodes in the merging environment. We evaluated the trained policy based on the merging vehicle's success rate defined by the completion of an episode without any collisions and the average reward value over the entire testing period.

\subsubsection*{Comparison and Analysis}

\begin{figure}[H]
\includegraphics[width=0.9\textwidth]{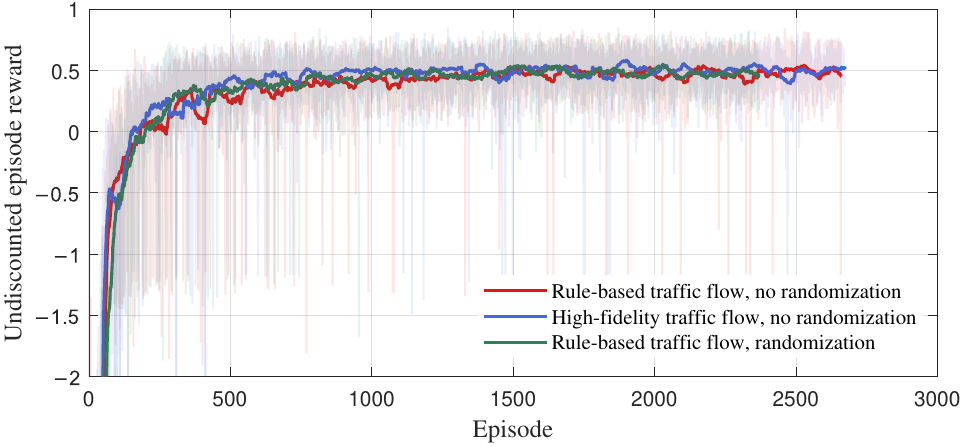}
\caption{Undiscounted 
 episode reward during training under three traffic flows.}
\label{f:merging}
\end{figure}

The training curves depicted in Figure \ref{f:merging} suggest that there is minimal visible difference in the rate of convergence and the rewards achieved by strategies trained under different microscopic traffic flows. 

Table \ref{t:merging} shows that the policy trained under rule-based traffic flow without randomization and high-fidelity microscopic traffic flow yields poor results when adapted to domain-randomized rule-based traffic flow. Conversely, the policy trained under domain-randomized rule-based traffic flows consistently achieves success rates above $90\%$ when tested across all three traffic flows.

\vspace{-12pt}
\begin{table}[H]
\caption{The results of testing the trained policies regarding merging.\label{t:merging}}
	\begin{adjustwidth}{-\extralength}{0cm}
		\begin{tabularx}{\fulllength}{CCCCCCC}
			\toprule
                \textbf{}	& \textbf{} & \multicolumn{4}{c}{\textbf{Traffic Flows for Training}} \\
                \cmidrule(l){3-6}
			\textbf{}	& \textbf{}	& \textbf{} 
   & \textbf{\shortstack{Rule-Based,\\No Randomization}} 
   & \textbf{\shortstack{High-Fidelity,\\No Randomization}} 
   & \textbf{\shortstack{Rule-Based,\\Randomization}}\\
			\midrule
\multirow[m]{6}{*}{\textbf{Testing 
}\vspace{-6pt}}& \multirow[m]{2}{*}{\textbf{\shortstack{rule-based,\\no randomization}}} & Average reward & 0.0058 & 0.0002 & $-$0.0018 
\\
			             & \textbf{} & Success rate & 98.50\% & 76.40\% & $91.60\%$\\
\cmidrule(l){2-6}
&\multirow[m]{2}{*}{\textbf{\shortstack{high-fidelity,\\no randomization}}}    & Average reward & 0.0029 & 0.0055 & $-$0.0069\\
			            & \textbf{} & Success rate & 95.20\% & 97.50\% & $98.20\%$\\
\cmidrule(l){2-6}
&\multirow[m]{2}{*}{\textbf{\shortstack{rule-based,\\randomization}}}    & Average reward & $-$0.0089 & $-$0.0065 & $0.0057$\\
			            & \textbf{} & Success rate & 56.00\% & 66.70\%\ & $99.30\%$\\
			\bottomrule
		\end{tabularx}
	\end{adjustwidth}
\end{table}

\subsubsection{Generalization Results for Increased Traffic Densities}

High-fidelity microscopic traffic flows closely resemble actual traffic scenes, so we used them as the test traffic flow with increased traffic densities. The impact of changes in traffic density is shown in Table \ref{t:merging_density}. It can be observed that the policy trained under rule-based traffic flow without randomization experiences a gradual decline in success rates and rewards as traffic density increases. In contrast, the policy trained under domain-randomized rule-based traffic flow consistently maintains a higher success rate.

\begin{table}[H]
\caption{The impact of traffic densities on three trained policies with high-fidelity traffic flow.\label{t:merging_density}}
	\begin{adjustwidth}{-\extralength}{0cm}
		\begin{tabularx}{\fulllength}{CCCCCCC}
			\toprule
            \textbf{}	&\textbf{}  & \multicolumn{4}{c}{\textbf{Traffic Density for Testing under High-Fidelity Traffic Flow}} \\
                \cmidrule(l){3-6}
			\textbf{}	& \textbf{}	& \textbf{} & $\boldsymbol{\phi=0.56}$    & $\boldsymbol{\phi=0.72}$  & $\boldsymbol{\phi=0.89}$\\
			\midrule
\multirow[m]{4}{*}{\textbf{\shortstack{Training under \\ \ rule-based traffic\\ flow~$(\boldsymbol{\phi=0.56})$}}\vspace{-6pt}}
&\multirow{2}{*}{\textbf{no randomization}}	
			  	        & Average reward & 0.0039 & 0.0017 & 0.0006\\
&	           & Success rate & 95.90\% & 93.30\% & 91.90\%\\
\cmidrule(l){2-6}
&\multirow{2}{*}{\textbf{randomization}}
	& Average reward & $-$0.0001 
 & $-$0.0002 & $-$0.0001\\
&		          & Success rate & 98.20\% & 98.50\% & 98.20\%\\
			\bottomrule
		\end{tabularx}
	\end{adjustwidth}
\noindent{\footnotesize{
 $\phi$ is the vehicle generation probability of the microscopic traffic flow, defined as the number of vehicles that are generated from the lane starting point per second.}}
\end{table}

\subsubsection{Ablation Study}\label{sec:ablation_study_freeway}
In order to strengthen the understanding of individual domain-randomized parameters' role in the model's performance, we analyzed their individual impact on the training outcomes through an ablation study. For the ablation study, we separately ablated each of the domain-randomized parameters. Subsequently, policies were individually trained under the traffic flows with domain-randomized parameter ablation. Finally, the trained policies were tested under both the domain-randomized (all parameters randomized) and high-fidelity traffic flows. The results of the ablation study are shown in Table~\ref{t:ablation_merging}. 

\begin{table}[H]
\caption{The 
 results of the ablation study.\label{t:ablation_merging}}
	\begin{adjustwidth}{-\extralength}{0cm}
		\begin{tabularx}{\fulllength}{CCCCCCC}
			\toprule
                \multirow[c]{2}{*}{\textbf{\shortstack{\\ \\ \\Training under\\Rule-Based Traffic Flow}}}	& \multicolumn{3}{c}{\textbf{Traffic Flows for Testing}} \\
                \cmidrule(l){2-4}
			&\textbf{} 
   & \textbf{\shortstack{Rule-Based,\\Randomization}} 
   & \textbf{\shortstack{High-Fidelity,\\No Randomization}} \\
			\midrule
\multirow[m]{2}{*}{\textbf{randomization---no
}~~$\boldsymbol{\delta}$} & Average reward & 0.0049 & $-$0.0023 \\
			             & Success rate & 99.90\% & 95.60\% \\
\midrule
\multirow[m]{2}{*}{\textbf{randomization---no}~~$\boldsymbol{T}$}    & Average reward & 0.0018 & $-$0.0049 \\
			            & Success rate & 92.50\% & 94.10\% \\
\midrule
\multirow[m]{2}{*}{\textbf{randomization---no}~~$\boldsymbol{a_{max}}$}    & Average reward & 0.0038 & $-$0.0038 \\
			            & Success rate & 97.30\% & 97.70\% \\
\midrule
\multirow[m]{2}{*}{\textbf{randomization---no}~~$\boldsymbol{a_{min}}$}    & Average reward & 0.0025 & $-$0.0026 \\
			            & Success rate & 94.20\% & 96.80\% \\
\midrule
\multirow[m]{2}{*}{\textbf{randomization---no}~~$\boldsymbol{v_{max}}$}    & Average reward & $-$0.0070 & 0.0016 \\
			            & Success rate & 65.00\% & 93.70\% \\
\midrule
\multirow[m]{2}{*}{\textbf{no randomization}}    & Average reward & $-$0.0089 & 0.0029 \\
			            & Success rate & 56.00\% & 95.20\% \\
\midrule
\multirow[m]{2}{*}{\textbf{randomization---all parameters}}    & Average reward & 0.0057 & $-$0.0069 \\
			            & Success rate & 99.30\% & 98.20\% \\
			\bottomrule
		\end{tabularx}
	\end{adjustwidth}
\end{table}

It can be observed that a decline occurs in the performance of the policies trained under the traffic flows with ablations when tested under the high-fidelity traffic flow. Moreover, the ablation of $v_{max}$ 
 significantly affects performance.

\subsection{Freeway}
\subsubsection{Freeway Environment}
\begin{figure}[H]
\includegraphics[width=1\textwidth]{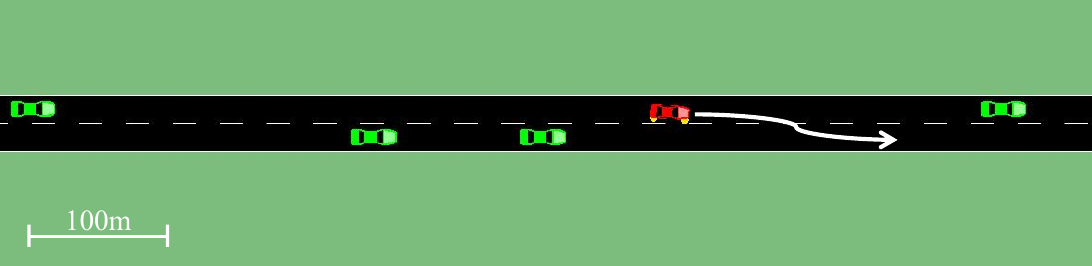}
\caption{The ego vehicle overtakes along the arrow trajectory in the freeway. 
}
\label{f:SUMO_freeway}
\end{figure}

\textls[-25]{We used a straight two-lane freeway measuring 1000 m in length, inspired by \mbox{Lin et al.~\cite{lin2024discretionary}}. Figure \ref{f:SUMO_freeway} depicts a standard lane-changing scenario in SUMO, where the ego vehicle is indicated by the red car and the surrounding vehicles are represented by the green cars. }
\subsubsection*{State}
The state of the environment is centered on the ego vehicle and four nearby vehicles: one directly in the front and one directly behind it in the same lane, and two similarly positioned vehicles in the adjacent lane. At time $t$, the state is defined by the longitudinal distance ($d^p_t,d^f_t,d^{adjacent_p}_t,d^{adjacent_f}_t$) of these four vehicles from the ego vehicle, their respective speeds ($v^p_t,v^f_t,v^{adjacent_p}_t,v^{adjacent_f}_t$), and the speed and acceleration ($v^{ego}_t,a^{ego}_t$) of the ego vehicle. These parameters form a state representation with ten variables as follows:
\begin{equation}\label{eq:freeway_s}
	s_t=(d^p_t,d^f_t,d^{adjacent_p}_t,d^{adjacent_f}_t,v^p_t,v^f_t,v^{adjacent_p}_t,v^{adjacent_f}_t,v^{ego}_t,a^{ego}_t)\in S.
\end{equation}
\subsubsection*{Action}
The action space is defined as follows:
\begin{equation}\label{eq:freeway_a}
	a_t = \{{acc^{ego}_t,0,1}\} \in A.
\end{equation}
where $acc^{ego}_t$ is a continuous action that indicates the acceleration of ego vehicle. Meanwhile, the discrete actions `0' and `1' dictate lane-changing behavior. The `0' means to keep the current lane and the `1' means an instantaneous lane change to the other lane.
\subsubsection*{Reward}
We have formulated a reward function aligned with practical driving objectives, incentivizing behaviors such as avoiding collisions, obeying speed limits, preserving comfortable driving conditions, and maintaining a safe following distance. The total reward $R_{\text{total}}$ is expressed as follows:
\begin{equation}
	R_{\text{total}} = R_{\text{act}}+R_{\text{distance}}+R_{\text{jerk}}+R_{v}+R_{\text{collision}}.
\end{equation}
In order to penalize frequent lane changes, the penalty $R_{\text{act}}$ is defined as follows: 
\begin{equation}\label{eq:Ract}
R_{\text{act}} = 
\begin{cases} 
  \omega_0, & \text{if } \left| y_{t-1} - y_t \right| \neq 0 \text{ and } d^p_t < d_{\text{safe}}, \\
  \omega_1, & \text{if } \left| y_{t-1} - y_t \right| \neq 0 \text{ and } d^p_t > d_{\text{safe}},  \\
  0, & \text{other},
\end{cases}
\end{equation}
where $y_t$ is ego vehicle's lateral position and $\omega_0$ < $\omega_1$. If the vehicle changes lanes within the safety distance $d_{\text{safe}}$, it incurs a penalty of $\omega_0$. Alternatively, changing lanes outside $d_{\text{safe}}$ results in a penalty of $\omega_1$.

It is essential to ensure that the ego vehicle maintains a safe following distance from the preceding vehicle, and the corresponding penalizing reward $R_{\text{distance}}$ is defined as
\begin{equation}\label{eq:freeway_Rd}
	R_{\text{distance}} = \omega_2 * \left| \displaystyle \frac{d^p_t - d_{\text{safe}}}{d_{\text{safe}}} \right|,  \text{if } d^p_t < d_{\text{safe}}.
\end{equation}
The objective of $R_{\text{jerk}}$ is to ensure driving comfort. It is defined as
\begin{equation}\label{eq:freeway_Racc}
	R_{\text{jerk}} = \omega_3 * \left| {(a_t - a_{t-1})}/0.1 \right|,
\end{equation}
where $a_t$ and $a_{t-1}$ denote the acceleration at the current and previous moments, respectively.

In order to promote the ego vehicle speed that enables overtaking, the penalty $R_{v}$ is defined as follows:
\begin{equation}\label{eq:freeway_Rv}
R_{v} = 
\begin{cases}
\omega_4 * \left| \frac{v^{ego}_t - v_{\text{stable}}}{v_{\text{safe}}} \right|, 
    & \text{if } v_{\text{stable}} < v^{ego}_t < v_{\text{safe}} \text{ and } d^p_t < d_{\text{safe}} + d^*,\\ \addlinespace
\omega_5 * \left| \frac{v^{ego}_t - v_{\text{safe}}}{v_{\text{safe}}} \right|, 
    & \text{if } v^{ego}_t > v_{\text{safe}}  \text{ and } d^p_t < d_{\text{safe}} + d^*,\\ \addlinespace
\omega_6 * \left| \frac{v^{ego}_t - v_{\text{stable}}}{v_{\text{stable}}} \right|,
    & \text{if } v^{ego}_t < v_{\text{stable}} \text{ and } d^p_t < d_{\text{safe}} + d^*,\\ \addlinespace
0, & \text{otherwise.}
\end{cases}
\end{equation}
When there is no opportunity to overtake the vehicle ahead, the ego vehicle should travel at a steady speed similar to that of the preceding vehicle. Consequently, we introduce a threshold $d^*$. As long as $d^p \in [d_{\text{safe}}, d_{\text{safe}} + d^*]$, the ego vehicle will not incur penalties of $R_{\text{distance}}$ or $R_{v}$.

In the equations presented above, $\omega_i$ denotes the corresponding weights. The key parameters for the freeway scene are presented in Table \ref{t:freeway_parameters}.

\begin{table}[H] 
	\caption{Parameters for freeway simulation.\label{t:freeway_parameters}}
	\begin{tabularx}{\textwidth}{CCCC}
		\toprule
		\textbf{Parameters}	& \textbf{Value}	& \textbf{Weights} & \textbf{Value}  \\
		\midrule
		$a_{\text{min}}$		&$-4.5$ m/s$^2$		& $w_0$   &$-5$\\
        $a_{\text{max}}$		&2.6 m/s$^2$		& $w_1$   &$-2$\\
		$v_{\text{safe}}$		& 16.89 m/s		& $w_2$   &$-10$\\
		$v_{\text{stable}}$	& 8.89 m/s	& $w_3$   &$-0.005$\\
		$d_{\text{safe}}$		& 25 m		    & $w_4$   &1\\
		$R_{\text{collision}}$		& $-200$		    & $w_5$       &$-0.5$\\
            $d^*$		& 2.5 m		    & $w_6$       &$-0.5$\\
		\bottomrule
	\end{tabularx}
\end{table}

\subsubsection{Parameterized Soft Actor--Critic}
The SAC algorithm in Section~\ref{subsec:sac} can only solve continuous-action space problems. When dealing with continuous-discrete hybrid action space for freeway lane change, we adopt the Parameterized SAC (PASAC) algorithm, inspired by Lin et al.~\cite{lin2024discretionary}. 

PASAC is based on SAC. The actor network produces continuous outputs, which include both continues actions and the weights for the discrete actions. An argmax function is utilized to select the discrete action associated with the maximum weight.

The freeway environment, having a hybrid continuous-discrete action space, requires the agent to be trained with the PASAC algorithm. The hyperparameters of PASAC are the same as SAC.

\subsubsection{Results Under Different Microscopic Traffic Flows}
\subsubsection*{Training}
In the freeway environment, we trained 400,000 timesteps in each of three different microscopic traffic flows. It required 1.5 h to complete the training using both rule-based traffic flows without randomization and with domain randomization. In contrast, the training under the condition of high-fidelity traffic flow took 5 h. Vehicle generation probability is 0.14 vehicles per second, and the traffic density on the main road was approximately 11 vehicles per kilometer on each straightaway.

\subsubsection*{Testing}
The trained policy was tested with 1000 episodes in the freeway environment. We evaluated the trained policy based on the ego vehicle's success rate defined by the completion of an episode without any collisions, and the average reward value over the entire testing~period.
\subsubsection*{Comparison and Analysis}

\begin{figure}[H]
\includegraphics[width=0.9\textwidth]{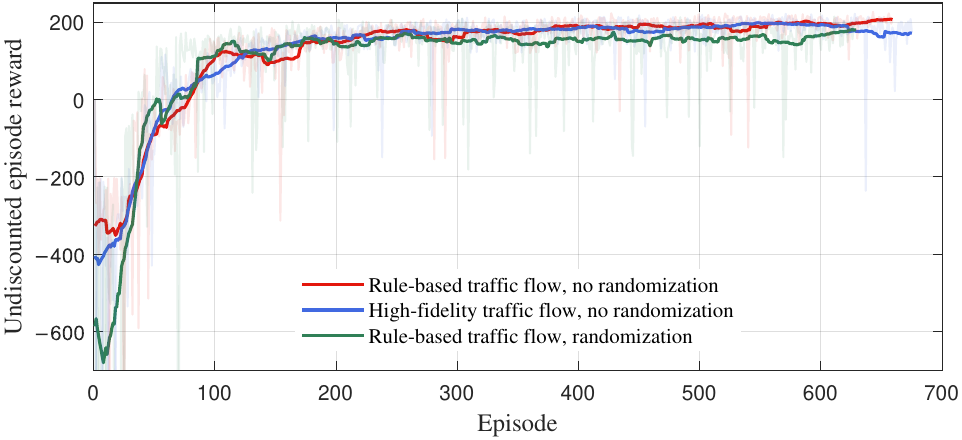}
\caption{Undiscounted 
 episode reward during training under three traffic flows.}
\label{f:freeway}
\end{figure}

In Figure \ref{f:freeway}, it can be observed that the policies all tend to converge around 200~episodes. Throughout the training process, aside from the initially lower reward of the domain-randomized traffic flows, the convergence rates and final rewards of the three curves are closely aligned.

The results of testing are shown in Table \ref{t：freeway}. It can be observed that the policy trained under domain-randomized rule-based traffic flows has the highest success rates when tested under different microscopic traffic flows. The policy trained under rule-based and high-fidelity traffic flows without randomization cannot adapt to domain-randomized rule-based traffic flow.
\vspace{-12pt}
\begin{table}[H]
\caption{The results of testing the trained policies regarding freeway condition.\label{t：freeway}}
	\begin{adjustwidth}{-\extralength}{0cm}
		\begin{tabularx}{\fulllength}{CCCCCCC}
			\toprule
                \textbf{}	& \textbf{} & \multicolumn{4}{c}{\textbf{\textbf{Traffic Flows for Training}}} \\
                \cmidrule(l){3-6}
			\textbf{}	& \textbf{}	& \textbf{} 
   & \textbf{\shortstack{Rule-Based,\\No Randomization}}
   & \textbf{\shortstack{High-Fidelity,\\No Randomization}} 
   & \textbf{\shortstack{Rule-Based,\\Randomization}}\\
			\midrule
\multirow[m]{6}{*}{\textbf{Testing 
}\vspace{-6pt}}& \multirow[m]{2}{*}{\textbf{\shortstack{rule-based,\\no randomization}}} & Average reward & 200.50 & 205.32 & 197.15 
\\
			             & \textbf{} & Success rate & 100\% & 99.70\% & $100\%$\\
\cmidrule(l){2-6}
&\multirow[m]{2}{*}{\textbf{\shortstack{high-fidelity,\\no randomization}}}     & Average reward & 187.10 & 208.85 & 202.48\\
			            & \textbf{} & Success rate & 98.90\% & 99.60\% & 99.90\%\\
\cmidrule(l){2-6}
&\multirow[m]{2}{*}{\textbf{\shortstack{rule-based,\\randomization}}}    & Average reward & 126.27 & 160.06 & 186.72\\
			            & \textbf{} & Success rate & 80.40\% & 89.00\%\ & 99.40\%\\
			\bottomrule
		\end{tabularx}
	\end{adjustwidth}
\end{table}

\subsubsection{Generalization Results for Increased Traffic Densities}

The impact of changes in high-fidelity traffic density is shown in Table \ref{t:density}. It can be observed that the success rate and reward of the policy trained under microscopic traffic flow without randomization decreases significantly as traffic density increases. In contrast, the policy trained under domain-randomized traffic flow maintains a success rate near~$100\%$.

\begin{table}[H]
\caption{The impact of changes in traffic density on three trained policies under high-fidelity traffic flow.\label{t:density}}
	\begin{adjustwidth}{-\extralength}{0cm}
		\begin{tabularx}{\fulllength}{CCCCCCC}
			\toprule
            \textbf{}	&\textbf{} & \multicolumn{4}{c}{\textbf{Traffic Density for Testing under High-Fidelity Traffic Flow}} \\
                \cmidrule(l){3-6}
			\textbf{}	& \textbf{}	&\textbf{} & $\boldsymbol{\phi=0.14}$    & $\boldsymbol{\phi=0.18}$  & $\boldsymbol{\phi=0.20}$\\
			\midrule
\multirow[m]{6}{*}{\textbf{\shortstack{Training under \\ \ rule-based traffic\\ flow $(\boldsymbol{\phi=0.14})$}}}
&\multirow{3}{*}{\textbf{no randomization}}	& Average speed	& 16.39	& 15.89	& 15.77 \\
&\textbf{}			  	        & Average reward & 187.10 & 189.85 & 109.49\\
&           & Success rate & 98.90\% & 93.90\% & 66.50\%\\
\cmidrule(l){2-6}
&\multirow{3}{*}{\textbf{randomization}} & Average speed & 15.83 & 15.42 & 15.17\\
&\textbf{}	& Average reward & 202.48 
 & 197.48 & 191.51\\
&	          & Success rate & 99.90\% & 99.80\% & 99.90\%\\
			\bottomrule
		\end{tabularx}
	\end{adjustwidth}
\noindent{\footnotesize{
 $\phi$ is the vehicle generation probability of the microscopic traffic flow, defined as the number of vehicles that are generated from the lane starting point per second.}}
\end{table}

\subsubsection{Ablation Study}
In the freeway environment, we also conducted an ablation study to enhance our understanding of the role that an individual domain-randomized parameter plays in the model's performance. The results of the ablation study are shown in Table~\ref{t:ablation_freeway}. 

\begin{table}[H]
\caption{The results of the ablation study.\label{t:ablation_freeway}}
	\begin{adjustwidth}{-\extralength}{0cm}
		\begin{tabularx}{\fulllength}{CCCCCCC}
			\toprule
                \multirow[c]{2}{*}{\textbf{\shortstack{\\ \\ \\Training under\\Rule-Based Traffic Flow}}}	& \multicolumn{3}{c}{\textbf{Traffic Flows for Testing}} \\
                \cmidrule(l){2-4}
			&\textbf{} 
   & \textbf{\shortstack{Rule-Based,\\Randomization}} 
   & \textbf{\shortstack{High-Fidelity,\\No Randomization}} \\
			\midrule
\multirow[m]{2}{*}{\textbf{randomization---no}~~$\bm{\delta}$} & Average reward & 170.46 & 187.65 \\
			             & Success rate & 97.20\% & 99.90\% \\
\midrule
\multirow[m]{2}{*}{\textbf{randomization---no}~~$\boldsymbol{T}$}    & Average reward & 169.62 & 178.95 \\
			            & Success rate & 98.50\% & 99.60\% \\
\midrule
\multirow[m]{2}{*}{\textbf{randomization---no}~~$\boldsymbol{a_{max}}$}    & Average reward & 181.49 & 202.52 \\
			            & Success rate & 98.30\% & 99.90\% \\
\midrule
\multirow[m]{2}{*}{\textbf{randomization---no}~~$\boldsymbol{a_{min}}$}    & Average reward & 170.72 & 185.81 \\
			            & Success rate & 99.40\% & 100\% \\
\midrule
\multirow[m]{2}{*}{\textbf{randomization---no}~~$\boldsymbol{v_{max}}$}    & Average reward & 96.99 & 150.10 \\
			            & Success rate & 89.90\% & 99.50\% \\
\midrule
\multirow[m]{2}{*}{\textbf{randomization---no}~\texttt{lcSpeedGain}}    & Average reward & 132.04 & 156.61 \\
			            & Success rate & 100\% & 100\% \\
\midrule
\multirow[m]{2}{*}{\textbf{randomization---no}~\texttt{lcAssertive}}    & Average reward & 173.71 & 188.57 \\
			            & Success rate & 97.70\% & 99.90\% \\
\midrule
\multirow[m]{2}{*}{\textbf{no randomization}}    & Average reward & 126.27 & 187.10 \\
			            & Success rate & 80.40\% & 98.90\% \\
\midrule
\multirow[m]{2}{*}{\textbf{randomization---all parameters}}    & Average reward & 186.72 & 202.48 \\
			            & Success rate & 99.40\% & 99.90\% \\
			\bottomrule
		\end{tabularx}
	\end{adjustwidth}
\end{table}

In the freeway environment, the collision rates of the different policies are close to zero, so we mainly compare the average rewards of the different policies. It can be found that the conclusions are similar to those of merging, where the performance of the policies trained under traffic flows with domain-randomized parameter ablation declines to varying degrees. The ablation of $v_{max}$ has a large impact on the performance. 

\section{Conclusions}\label{sec:Conclusion}
In this study, we introduce a method for randomizing lane-changing and car-following model parameters to generate randomized microscopic traffic flows, and we evaluate and compare the policies trained by reinforcement learning algorithms in freeway and merging environments. The results show that
\begin{itemize}
\item The policy trained under the condition of domain-randomized rule-based microscopic traffic flow is able to maintain  high rewards and success rates when tested with different microscopic traffic flows. However, the policy trained under the condition of microscopic traffic flow without randomization or high-fidelity microtraffic flow performs significantly worse when tested under microscopic traffic flows that are different from those of training. This indicates that domain randomization enables reinforcement learning agents to adapt to different types of traffic flow.

\item The policy trained under the condition of domain-randomized rule-based microscopic traffic flow performs well when tested under high-fidelity microscopic traffic flow with different traffic densities. The policy trained under microscopic traffic flow without randomization decreases significantly with increasing traffic density. This indicates that the domain-randomized traffic flow possesses strong generalization to changes in traffic~density.

\item Although high-fidelity microscopic traffic flow is close to real microscopic traffic flows, the results show that not only does it considerably increase simulation time but policies trained under the condition of microscopic traffic flow also do not generalize well to different microscopic flows. Therefore, high-fidelity microscopic traffic flow is more suitable for testing rather than training.
\end{itemize}

In summary, the policies trained under domain-randomized rule-based microscopic  traffic flow demonstrate robust performance when transferred to environments that closely resemble real-world traffic conditions. The future work includes testing a policy trained under the condition of domain-randomized rule-based microscopic traffic flow on a real autonomous vehicle.

\vspace{6pt} 


\authorcontributions{Conceptualization, Y.L.; methodology, A.X., Y.L., and X.L.; formal analysis, A.X. and Y.L.; investigation, A.X. and Y.L.; data curation, A.X.; writing---original draft preparation, A.X.; writing---review and editing, A.X.; supervision, Y.L. All authors have read and agreed to the published version of the manuscript.}

\funding{This work was supported in part by Guangzhou Basic and Applied Basic Research Program under Grant 2023A04J1688, and in part by South China University of Technology faculty start-up~fund.}

\institutionalreview{Not applicable.}

\informedconsent{Not applicable.}

\dataavailability{The data can be obtained upon reasonable request from the corresponding author.} 

\conflictsofinterest{The authors declare that they have no known competing financial interests or
	personal relationships that could have appeared to influence the work reported in this paper.} 


\begin{adjustwidth}{-\extralength}{0cm}
\reftitle{References}

\PublishersNote{}
\end{adjustwidth}
\end{document}